\documentclass[aps,prl,floatfix,epsfig,twocolumn,showpacs,preprintnumbers]{revtex4}
\usepackage{amsmath,amssymb,graphicx,bm,epsfig}
\usepackage{color}
\usepackage{braket}

\renewcommand{\vr}{{\mathbf{r}}}

\newcommand{\vk}{{\mathbf{k}}}

\newcommand{\Tr}{\mathrm{Tr}}

\begin{document}

\title{Exact double-counting in combining the Dynamical Mean Field
  Theory and the Density Functional Theory}

\author{Kristjan Haule}
\affiliation{Department of Physics and Astronomy, Rutgers University, Piscataway, USA}
\date{today}

\begin{abstract}
  We propose a continuum representation of the Dynamical Mean Field
  Theory, in which we were able to derive an exact overlap between the
  Dynamical Mean Field Theory and band structure methods, such as the
  Density Functional Theory.  The implementation of this exact
  double-counting shows improved agreement between theory and
  experiment in several correlated solids, such as the transition
  metal oxides and lanthanides. Previously introduced nominal
  double-counting is in much better agreement with the exact
  double-counting than most widely used fully localized limit formula.
\end{abstract}

\date{\today}
\maketitle

Understanding the electronic structure of materials with strong
electronic correlations remains one of the great challenges of modern
condensed matter physics.  The first step towards calculating the
electronic structure of solids has been achieved by obtaining the
single-particle band dispersion $E(\vk)$ within the density functional
theory (DFT) in the local density approximation (LDA)~\cite{DFT},
which takes into account correlation effects only in a limited extent.

To account for the many-body correlation effects beyond LDA, more
sophisticated methods have been developed. Among them, one of the most
successful schemes is the dynamical mean-field theory
(DMFT)~\cite{DMFT-first}. It replaces the problem of describing
correlation effects in a periodic lattice by a strongly interacting
impurity coupled to a self-consistent bath~\cite{DMFT-RMP1996}. This
method was first developed to solve the Hubbard model, but it was soon
realized~\cite{Gabi-first} that it can also be combined with the LDA
method, to give more material specific predictions of correlation
effects in solids. The LDA+DMFT method achieved great success in the
past two decades, as it was successfully applied to numerous
correlated solids~\cite{review}.  The combination of the two methods,
nevertheless lead to a problem of somewhat ambiguous way of
subtracting the part of correlations which are accounted for by both
methods.

The so-called double-counting (DC) term was usually approximated by
the formula first developed in the context of LDA+U, and was evaluated
by taking the atomic limit for the Hubbard interaction
term~\cite{Sawatzky,Anisimov}. Many other similar schemes were
proposed recently~\cite{LichtensteinNiO,Haule-DMFT,covalency,Park},
but rigorous derivation of this double-counted interaction within DMFT
and LDA is missing to date. Here we propose a new method of
calculating the overlap between DMFT and a band-structure method
(either DFT or GW), and we explicitly evaluate this DC functional
within LDA+DMFT applied to well studied transition metal oxides such
as SrVO$_3$, LaVO$_3$, and most studied lanthanide metal, the
elemental Cerium.

To compare different approximations in the same language, it is useful
to cast them into the form of the Luttinger Ward
functional~\cite{Luttinger-Ward,Baym-Kadanoff,review}, which is a
functional of the electron Green's function $G$, and takes the form
$\Gamma[G]=-\Tr((G_0^{-1}-G^{-1})G)+\Tr\log(-G)+\Phi_{V_c}[G]$.  The
first part is the material dependent part, in which
$G_0^{-1}(\vr\vr';\omega)=[\omega+\mu+\nabla^2-V_{ext}(\vr)]\delta(\vr-\vr')$,
and the second two terms are universal functionals of the Green's
function $G(\vr\tau,\vr'\tau')$ and the Coulomb interaction
$V_c(\vr-\vr')$. In the exact theory, $\Phi_{V_c}[G]$ contains all
skeleton Feynman diagram, constructed by $G$ and
$V_c$~\cite{Baym-Kadanoff}.  In the language of Luttinger Ward
functional, different approximations can then be looked at as
different approximations to the interacting part $\Phi_{V_c}[G]$.

The Density Functional Theory can be derived by approximating the
exact functional $\Phi_{V_c}[G]$ by $E_H[\rho(\vr)] +
E_{xc}[\rho(\vr)]$, where $E_H$ and $E_{xc}$ are the Hartree and
the exchange-correlation functionals, respectively. The stationarity condition gives the DFT
equations, i.e.,
$G^{-1}-G_0^{-1}=(V_H[\rho]+V_{xc}[\rho])\delta(\vr-\vr')\delta(\tau-\tau')$,
because $\delta E_{xc}[\rho]/\delta G =
\delta(\vr-\vr')\delta(\tau-\tau')\; \delta
E_{xc}[\rho]/\delta\rho=\delta(\vr-\vr')\delta(\tau-\tau')\;
V_{xc}[\rho]$.  Note that in this language, exact DFT appears as an
approximation to the exact Green's function, where the exact
self-energy is approximated by a static and local potential. Note also
that the static approximation is a consequence of truncating the
variable of interest, namely replacing full $G(\vr,\tau,\vr',\tau')$
by its diagonal components
$\rho(\vr)=\delta(\vr-\vr')\delta(\tau-\tau')G(\vr\tau,\vr'\tau')$.

In the Luttinger-Ward functional language, the DMFT appears as an
approximation where the Green's function in the $\Phi$ functional is
replaced by its local counterpart $G\rightarrow G_{local}$ , and the
Coulomb repulsion $V_c$ by screened interaction $V_c\rightarrow U$,
namely $\Phi^{DMFT}=\Phi_{U}[G_{local}]$.~\cite{review} Note that the
DMFT functional has exactly the same form as the exact functional
$\Phi_{Vc}[G]$, because all the skeleton Feynman diagrams constructed
by $G_{local}$ and $U$ are summed up by DMFT, while in DFT the
functional $E_{xc}[\rho]$ is unknown, and further approximation is
necessary. The truncation of the variable of interest from $G$ to
$G_{local}$ leads in DMFT to self-energy, which is also local in
space, but it keeps its dynamic nature. Other approximations such as
Hartree-Fock or GW can be similarly derived by replacing
$\Phi_{Vc}[G]$ by some limited set of Feynman diagrams, i.e.,
truncation in space of Feynman diagrams, rather than truncation of the
variable of interest.

There is some kind of disconnect between the DMFT functional
$\Phi^{DMFT}_U[G_{local}]$, and the LDA functional
$E_{xc}[\rho(\vr)]$, mostly because the auxiliary systems for the two
methods are very different. The auxiliary system for LDA approximation
is the uniform electron gas problem defined for continuum, in the
absence of complexity of the solid. On the other hand, DMFT is usually
associated with the lattice model like Hubbard model, where mapping to
the local problem reduces to the Anderson impurity model, which does
not have a well-defined continuum representation.  The double-counting
problem occurs because it is not clear what is the overlap between the
two methods, i.e., what physical processes are accounted for in one
and what in the other method.

It is useful to represent the DMFT method in the continuum
representation. Such representation is not unique, but physical
intuition can guide the mapping. Here we propose to look at the DMFT
problem as the approximation, which solves exactly the problem defined
by some auxiliary Green's function $G_{local}=\hat{P} G$ and Coulomb
repulsion replaced by Yukawa short-range interaction $V_c^\lambda =
\frac{e^{-\lambda |\vr-\vr'|}}{|\vr-\vr'|}$. We have in mind some
projector $\hat{P}$, which is very local, and truncates the Green's
function to a region mostly concentrated inside the muffin-tin
sphere. It can for example be defined by a set of quasi-atomic
orbitals
$G_{local}(\vr,\vr')=\sum_{L,L'}\braket{\vr|\Phi_L}\bra{\Phi_L}G\ket{\phi_{L'}}\braket{\phi_{L'}|\vr'}$
where $\braket{\vr|\Phi_L}=u_l(r)Y_{L}(\vr)$ are spheric harmonics
times localized radial wave function.  Note that this truncation of
the Green's function $G(\vr,\vr')$ to its local counterpart parallels
the truncation of the Green's function to its diagonal component in
theories that choose density as the essential variable, i.e.,
$\rho(\vr)=G(\vr\tau,\vr'\tau')\delta(\vr-\vr')\delta(\tau-\tau')$.
The screening $\lambda$ in Yukawa interaction $V_c^\lambda$ has to be
large enough such that the interaction between electrons on
neighboring sites is negligible. The DMFT can then give an exact
Luttinger-Ward functional $\Phi_{V_c^\lambda}[\hat{P} G]$, i.e.,
containing all local Feynman diagrams constructed by $\hat{P} G$ and
$V_c^\lambda$, defined in the continuum~\cite{Chitra}.  The
stationarity condition for the Luttinger-Ward functional gives the
DMFT equations $G^{-1}-G_0^{-1}=\hat{P}\;
\left(\delta\Phi_{Vc^\lambda}[G_{local}]/\delta G_{local}\right)$.

The precise determination of the screening $\lambda$ is beyond the
scope of this paper. However, we notice that once the Coulomb
interaction $U$ in DMFT is known, the screening length $\lambda$ is
uniquely determined by $U$ through the matrix elements of the Yukawa
interaction in DMFT basis. Notice that Hund's coupling $J$ is not a
free parameter in this parametrization, but is uniquely determined by
$\lambda$ through Yukawa form of the Coulomb interaction.~\cite{online}

After the mapping of the DMFT method to the continuous ($\vr,\vr'$)
Hilbert space, where DFT exchange-correlation is defined, it is easy
to see what is the overlap between the two methods. The Hartree term
is accounted for exactly in the LDA method, and has the form
$E^H_{V_c}[\rho]=\frac{1}{2}\int d\vr d\vr' \rho(\vr)\rho(\vr')
V_c(\vr-\vr')$, while in DMFT it takes the following form $E^{H,DMFT}
= \frac{1}{2}\int d\vr d\vr' (\hat{P} \rho(\vr)) (\hat{P} \rho(\vr'))
V_c^\lambda(\vr-\vr')$, which can also be written as $E^{H,DMFT} =
E^H_{V_c^\lambda}[\hat{P}\rho]$, where $\hat{P}\rho=
\delta(\vr-\vr')\delta(\tau-\tau')
G_{local}(\vr\tau,\vr'\tau')=\delta(\vr-\vr')\delta(\tau-\tau')\hat{P}G(\vr\tau,\vr'\tau')$,
and $E^H_{V_c}[\rho]$ is the exact Hartree functional defined
above. The Hartree contribution to the DC within LDA+DMFT (or any
other band structure method which includes exact Hartree term) is thus
$E^H_{V_c^\lambda}[\hat{P}\rho]$~\cite{Juho}. This DC term thus
corresponds to truncating Green's function $G$ and Coulomb interaction
$V_c$ by their local counterparts, i.e., $G\rightarrow \hat{P} G$ and
$V_c\rightarrow V_c^\lambda$.

For approximations, which truncate in the space of Feynman diagrams
(such as Hartree-Fock or GW method), one can obtain the DMFT
double-counting by applying both the truncation in space of Feynman
diagrams as well as the DMFT truncation in variables of interest. For
the case of GW method, one can check diagram by diagram that the
corresponding DMFT Feynman diagram is obtained by replacing $G$ by
$\hat{P} G$ and $V_c$ by $V_c^\lambda$ in each diagram, just like it
was done above for the Hartree term. More precisely, the GW functional
can be written as
$\Phi^{GW}_{V_c}[G]=E^{H}_{V_c}-\frac{1}{2}\Tr\log(1-V_c G*G)$, where
$G*G=P$ is the convolution of two Green's functions (polarization
function). The GW+DMFT double-counting is thus
$E^{H,DMFT}-\frac{1}{2}\Tr\log(1-V^\lambda_c (\hat{P}G)*(\hat{P}G))$,
which can be shortly written as $\Phi^{GW}_{V_c^\lambda}[\hat{P}G]$.

In the case of DFT+DMFT, the expansion in terms of Feynman diagrams is
not possible, however, to identify the overlap between the two
methods, this is not essential. Clearly, the double-counting in
DFT+DMFT is obtained by the same procedure of replacing $G$ by
$\hat{P}G$ and $V_{c}$ by $V_{c}^\lambda$ in the DFT functional. Since
the DFT also truncates the Green's function to its diagonal components
only ($\rho=\delta(\tau-\tau')\delta(\vr-\vr')G$) the DC is a
functional of the local charge only $\rho_{local}=\hat{P}\rho$. DC
thus takes the form
$$\Phi_{DC}^{DFT+DMFT} = E^H_{V_c^\lambda}[\hat{P}\rho] + E^{XC}_{V_c^\lambda}[\hat{P}\rho].$$
In LDA method, the exchange-correlation functional is obtained from
the energy of the uniform electron gas. To obtain the LDA+DMFT
double-counting, one thus needs to solve the problem of the electron
gas with the density that contains only "local'' charge $\hat{P}\rho$,
which interacts by the screened Yukawa interaction $V_{c}^\lambda$.~\cite{online}

Including the exact double-counting, the LDA+DMFT $\Phi$ functional is thus
\begin{eqnarray}
\Phi^{LDA+DMFT}[G]=E^H_{V_c}[\rho]+E^{XC}_{V_c}[\rho]+\Phi_{V_c^\lambda}[\hat{P}G]-\nonumber\\
-E^H_{V_c^\lambda}[\hat{P}\rho]-E^{XC}_{V_c^\lambda}[\hat{P}\rho],
\label{Eq:functional}
\end{eqnarray}
where $\Phi_{V_c^\lambda}[\hat{P}G]$ is the DMFT functional which
contains all Feynman diagrams constructed from $\hat{P}G$ and
$V_c^\lambda$.  This is the central equation of this paper, as it
defines the LDA+DMFT approximation including the exact DC.
The saddle point equations give the LDA+DMFT set of equations
\begin{eqnarray}
&&G^{-1}-G_0^{-1}=
\hat{P}\frac{\delta \Phi_{V_c^\lambda}[G_{local}]}{\delta G_{local}}
+\\
&&\left(
\frac{\delta  E^{HXC}_{V_c^\lambda}[\rho]}{\delta \rho}-
\hat{P}\frac{\delta  E^{HXC}_{V_c^\lambda}[\rho_{local}]}{\delta \rho_{local}}
\right)\delta(\vr-\vr')\delta(\tau-\tau')
\nonumber
\end{eqnarray}
where we used
$E^{HXC}[\rho]\equiv E^H[\rho]+E^{XC}[\rho]$ and $\hat{P}G\equiv G_{local}$.

The only difference between functional Eq.~\ref{Eq:functional}, and
the usual LDA+DMFT implementation, is the presence of
$E^{HXC}_{V_c^\lambda}$. This is the semi-local exchange and LDA
correlation functional of the electron gas interacting by Yukawa
interaction.
The semi-local exchange-density $\varepsilon^x_{V_c^\lambda}[\rho]$
(defined by $E^{x}[\rho]=\int d\vr
\rho(\vr)\varepsilon^x[\rho(\vr)]$), can be computed analytically, and
takes the following form
$$\varepsilon^x_{V_c^\lambda}[\rho]=-\frac{C}{r_s} f(x)$$
where
\begin{eqnarray}
f(x) =1-\frac{1}{6x^2}-\frac{4 \arctan(2x)}{3 x}+\frac{(12 x^2+1) \log(1+4x^2)}{24 x^4},
\nonumber
\end{eqnarray}
$C=\frac{3}{2}\left(\frac{9}{4\pi^2}\right)^{1/3}$, 
$r_s=\left(\frac{3}{4\pi\rho}\right)^{1/3}$, and
$x=\left(\frac{9\pi}{4}\right)^{1/3}\frac{1}{\lambda r_s}$.
The exchange potential
$V^x=\frac{\delta}{\delta\rho}E^x[\rho]$ is then
$V^x_{V_c^\lambda}=\frac{4}{3}\varepsilon^x_{V_c^\lambda}+\frac{1}{3}\frac{C}{r_s}
x\frac{df}{dx}$.

The correlation part requires solution of the homogeneous electron gas
problem interacting with Yukawa repulsion, which was solved by
QMC~\cite{Ceperly-Yukawa,Ceperly-QMC,Ortiz}.  Here we want to have an
analytic expression for correlation energy at arbitrary $\lambda$ and
$r_s$. It is well established that G$_0$W$_0$ gives quite accurate
correlation energy of the electron gas~\cite{GW-energy,GW-FSC},
especially when computed from the Luttinger-Ward functional
$\Gamma[G]$.  We thus repeated $G_0W_0$ calculation for the electron
gas, but here we use Yukawa interaction. We evaluate the total energy
using Luttinger-Ward functional of GW to achieve high accuracy. We
then fit the correlation energy in the range of physically most
relevant $r_s \in [0,10]$ and screenings $\lambda\in[0,3]$ ($\lambda$
is measured in Bohr radius inverse) with the following functional
form:
\begin{eqnarray}
\varepsilon^c_{V_c^\lambda} =
\frac{\varepsilon^c_{\lambda=0}}{1+\sum_{n=1}^4 a_n r_s^n}
\end{eqnarray}
where
\begin{eqnarray}
&&\log(1+a_1)=\frac{\lambda (\alpha_0+\alpha_1\lambda)}{1+\alpha_2\lambda^2+\alpha^3\lambda^4+\alpha_4\lambda^6}\\
&&\log(1+a_2)=\frac{\lambda^2(\beta_0+\beta_1\lambda)}{1+\beta_2\lambda^2+\beta_3\lambda^4}\\
&&\log(1+a_3)=\frac{\lambda^3(\gamma_0+\gamma_1\lambda)}{1+\gamma_2\lambda^2}\\
&&\log(1+a_4)=\lambda^4(\delta_0+\delta_1\lambda^2)
\end{eqnarray}
The best fit gives the following coefficients:
\begin{eqnarray}
&& \alpha_i=[1.2238912,  7.3648662,  9.6044695,\nonumber\\
&&\qquad\qquad\qquad        -0.7501634, 0.0207808]\times 10^{-1}\nonumber\\
&& \beta_i = [5.839362,  11.969474,  10.156124,  1.594125]\times 10^{-2}\nonumber\\
&& \gamma_i = [8.27519,  5.57133,  17.25079]\times 10^{-3}\nonumber\\
&& \delta_i = [5.29134419, 0.0449628225]\times 10^{-4}
\end{eqnarray}
Finally, the correlation potential is
$V^c_{\lambda}=\frac{V^c_{\lambda=0}}{A(r_s,\lambda)} +
\frac{\varepsilon^c_{\lambda=0}}{C(r_s,\lambda)}$, where
$A(r_s,\lambda)=1+\sum_{n=1}^4 a_n r_s^n$ and
$C(r_s,\lambda)=3[1+\sum_{n=1}^4 a_n r_s^n]^2/\sum_{n=1}^4 n\; a_n\;
r_s^n$.  We take the unscreened correlation energy density
$\varepsilon^c_{\lambda=0}$ (and unscreened potential) from the
standard parametrization of quantum Monte Carlo results, hence the
G$_0$W$_0$ calculation is only used for renormalization of
correlations by screening with Yukawa form.

In the following we present results for some of the most often studied
correlated solids, namely, elemental Cerium, SrVO$_3$ and LaVO$_3$.
We used three different forms of DC functional: i) "exact'', which we
introduced above, ii) "FLL'' stands for fully localized limit form
introduced in Ref.~\onlinecite{Sawatzky}, which has the simple form
$V_{dc}=U(n-1/2)-J/2(n-1)$, and $n$ stands for the correlated
occupancy, c) and the "nominal'' DC, introduced in
Ref.~\onlinecite{Haule-DMFT,covalency}.  The "nominal'' $V_{dc}$
takes the same form as "FLL'' formula, but $n$ in the formula is
replaced by the closest integer value ($n^0=[n]$), and hence $n^0$
corresponds to so-called nominal valence.  We use LDA+DMFT
implementation of Ref.~\onlinecite{Haule-DMFT}.

\begin {table}
\begin{center}
\begin{tabular}[c]{|l|c|c|}
\hline  
Ce-$\alpha$ & $n_f$ & $V_{dc}/U$ \\
\hline  
exact    &   0.997  &  0.424  \\
\hline                      
nominal&   1.002  &  0.500 \\
\hline                      
FLL        &  1.035   &  0.533 \\
\hline
\end{tabular}
\caption{LDA+DMFT valence and DC potential for $\alpha$-Ce at $T=200\,$K. The local Coulomb
  repulsion in Ce is $U=6\,$eV.
}
\label{Tabel1}
\end{center}
\end {table}
The physical properties of correlated materials are very sensitive to
the value of the local occupancy $n_f$, and $n_f$ is sensitive to the
value of DC.
In table~\ref{Tabel1} we show results for elemental Cerium 
in the $\alpha$ phase. All three DC functionals give very similar
correlated occupancies $n_f$, and all are very close to nominal
valence $n^0=1$. The actual value of the DC potential $V_{dc}$ differs
for less than $0.1\,U$, which leads to almost indistinguishable
spectra on the real axis, and from the previously published
results~\cite{Haule-DMFT}, hence we do not reproduce them here. We
found a general trend in all materials studied that the exact DC is
somewhat smaller then given by FLL formula.  For Ce, the Hartree
contribution to DC potential is $V_H = n_f U \approx 0.997\, U$, the
semi-local exchange contribution is $V_x \approx -0.485\, U$ and LDA
correlation is $V_c \approx -0.088\, U$, hence the total DC potential
is $V_H+V_x+V_c\approx 0.424 U$, which is slightly smaller than FLL
formula $U(n_f-1/2)-J/2(n_f-1)\approx 0.533 U$ or nominal formula
$U(n_f^0-1/2)-J/2(n_f^0-1)=0.5U$.  It is interesting to note that the
semi-local exchange used in LDA is quite different from the exact
exchange value.  The latter is only $|V_F| = U n/14\approx 0.071\, U$,
a substantially smaller value then the semi-local exchange $|V_x|
\approx 0.485\, U$.  This shows why DC within LDA+DMFT is so different
from the Hartree-Fock value of the DMFT self-energy, i.e.,
$\Sigma(\omega=\infty)$.

\begin {table}
\begin{center}
\begin{tabular}[c]{|l|c|c|c|c|c|}
\hline  
SrVO3 & $n_{t2g+eg}$ &$n_{t2g}$ & $n_{eg}$ & $V^{t2g}_{dc}/U$ & $V_{dc}^{eg}/U$ \\
\hline  
exact    & 2.223   & 1.507 & 0.716  & 1.384 &  1.406   \\
\hline                      
nominal&  2.251  & 1.541 & 0.710   & 1.443       & 1.444 \\
\hline                      
FLL        & 2.529   & 1.699  & 0.830   & 1.943   & 1.943 \\
\hline
\end{tabular}
\caption{LDA+DMFT results for SrVO$_3$ at $T=200\,$K and $U=10\,$eV. 
  Both $t2g$ and $eg$ orbitals are treated by DMFT.
}
\label{Tabel2}
\end{center}
\end {table}
\begin{figure}
\includegraphics[width=0.9\linewidth]{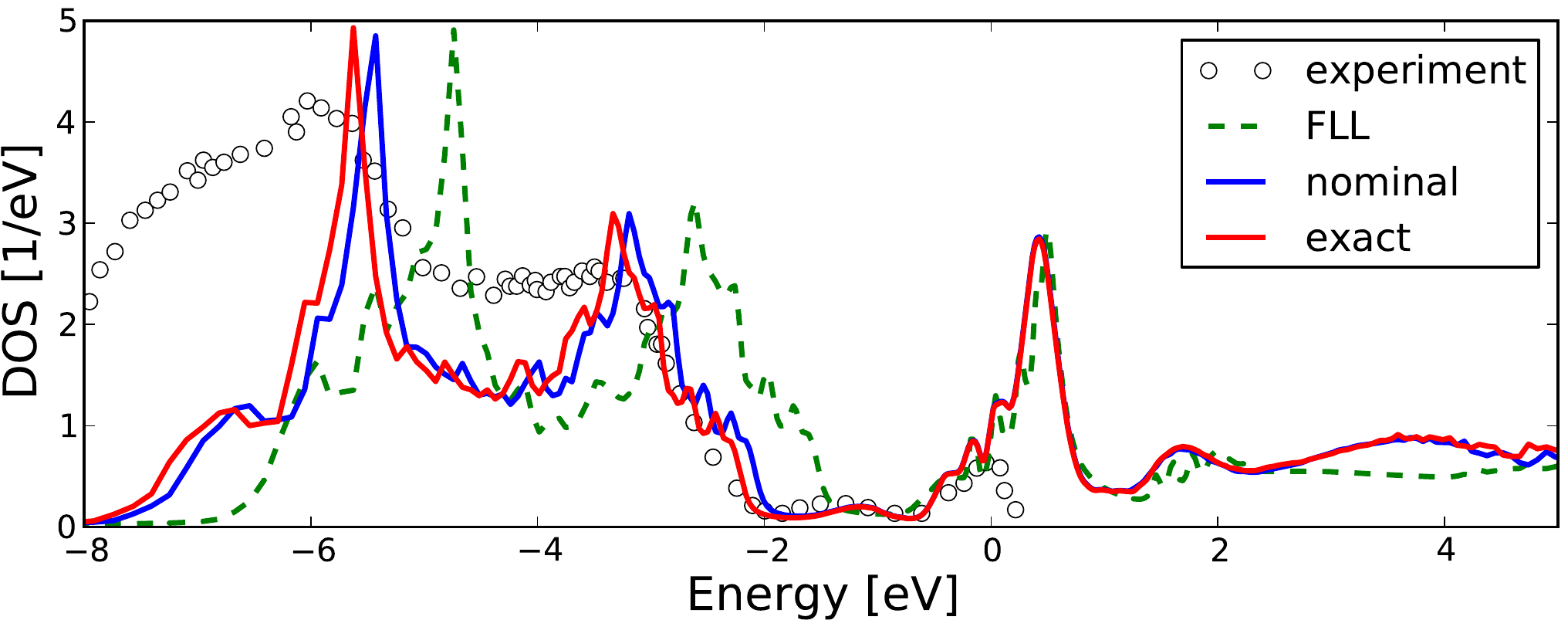}
\caption{(Color online) LDA+DMFT total density of states for
  SrVO$_3$ using three different DC potentials. Experimental
  photoemission is reproduced from Ref.~\onlinecite{SVO-PES}.
(parameters listed in table~\ref{Tabel2}).
}
\label{fig:srvo3}
\end{figure}
Next we present tests for SrVO$_3$, which is a metallic transition
metal oxide with nominally single electron in the t2g shell. Near the
Fermi level $E_F$, there are mostly $t2g$ states.  The majority of
$eg$ states are above $E_F$, however, due to strong hybridization with
oxygen some part of $eg$ orbitals also gets filled. There are two ways
the DMFT method can be used here. In the first case, one can treat
only the $t2g$ shell within DMFT. The vast majority of DMFT
calculations for SrVO3 were done in this way. In this case, all three
DC potentials again give very similar results and the spectra is
almost indistinguishable from previously published results in
Ref.~\onlinecite{covalency}.  One can also treat dynamically with DMFT
the entire $d$ shell.  This case is presented in Table~\ref{Tabel2}
and spectra in Fig.~\ref{fig:srvo3}. One can notice that the exact and
the nominal DC give very similar $n_d$, while the FLL formula gives
14\% larger $n_d$.
This is because the value of the DC potential is
substantially larger ($\approx 40\,$\%) when using FLL as compared to
exact case. It is nevertheless
comforting to see that 40\% error in double-counting still does not
leads to major failure of LDA+DMFT. We plot the spectra in
Fig.~\ref{fig:srvo3}, to show how this change in $V_{dc}$ leads to
shift of oxygen-$p$ spectra relative to vanadium-$d$ states. For the
exact DC, the oxygen peak positions match very well with the
experimentally measured spectra. The nominal valence is quite close to
the exact spectra, while FLL formula leads to an upward shift of
oxygen for roughly $0.6\,$eV, which is still relatively small compared
to the difference in the double-counting potentials, which is
$V_{dc}^{FLL}-V_{dc}^{exact}\approx 5.37\,$eV.

\begin {table}
\begin{center}
\begin{tabular}[c]{|l|c|c|c|c|c|}
\hline  
LaVO3(t2g-only) &$n_{t2g}$ & $V^{a1g}_{dc}/U$ & $V_{dc}^{eg'}/U$ \\
\hline  
exact    & 2.014 &  1.195  & 1.193 \\
\hline             
nominal& 2.074  & 1.450       &  1.450\\
\hline             
FLL       & 2.099  & 1.544  &  1.544\\
\hline
\end{tabular}
\caption{
LDA+DMFT results for LaVO$_3$ at $T=200\,$K and $U=10\,$eV. Only $t2g$
orbitals are treated by DMFT.
}
\label{Tabel3}
\end{center}
\end {table}
\begin {table}
\begin{center}
\begin{tabular}[c]{|l|c|c|c|c|c|c|}
\hline  
LaVO3(t2g+eg) & $n_{t2g+eg}$ &$n_{t2g}$ & $n_{eg}$ & $V^{a1g}_{dc}/U$& $V_{dc}^{eg'}/U$ & $V_{dc}^{eg}$\\
\hline  
exact    & 2.444  & 2.048 &  0.397 &   1.596  & 1.599 & 1.665\\
\hline                      
nominal& 2.344   & 2.032  & 0.312  &  1.458   & 1.458 & 1.458\\
\hline                      
FLL       & 2.706    & 2.167   & 0.540   & 2.114   & 2.114    & 2.114\\
\hline
\end{tabular}
\caption{
LDA+DMFT results for LaVO$_3$ at $T=200\,$K and $U=10\,$eV. Both $t2g$
and $eg$ orbitals are treated by DMFT.
}
\label{Tabel4}
\end{center}
\end {table}
\begin{figure}[htb]
\includegraphics[width=0.9\linewidth]{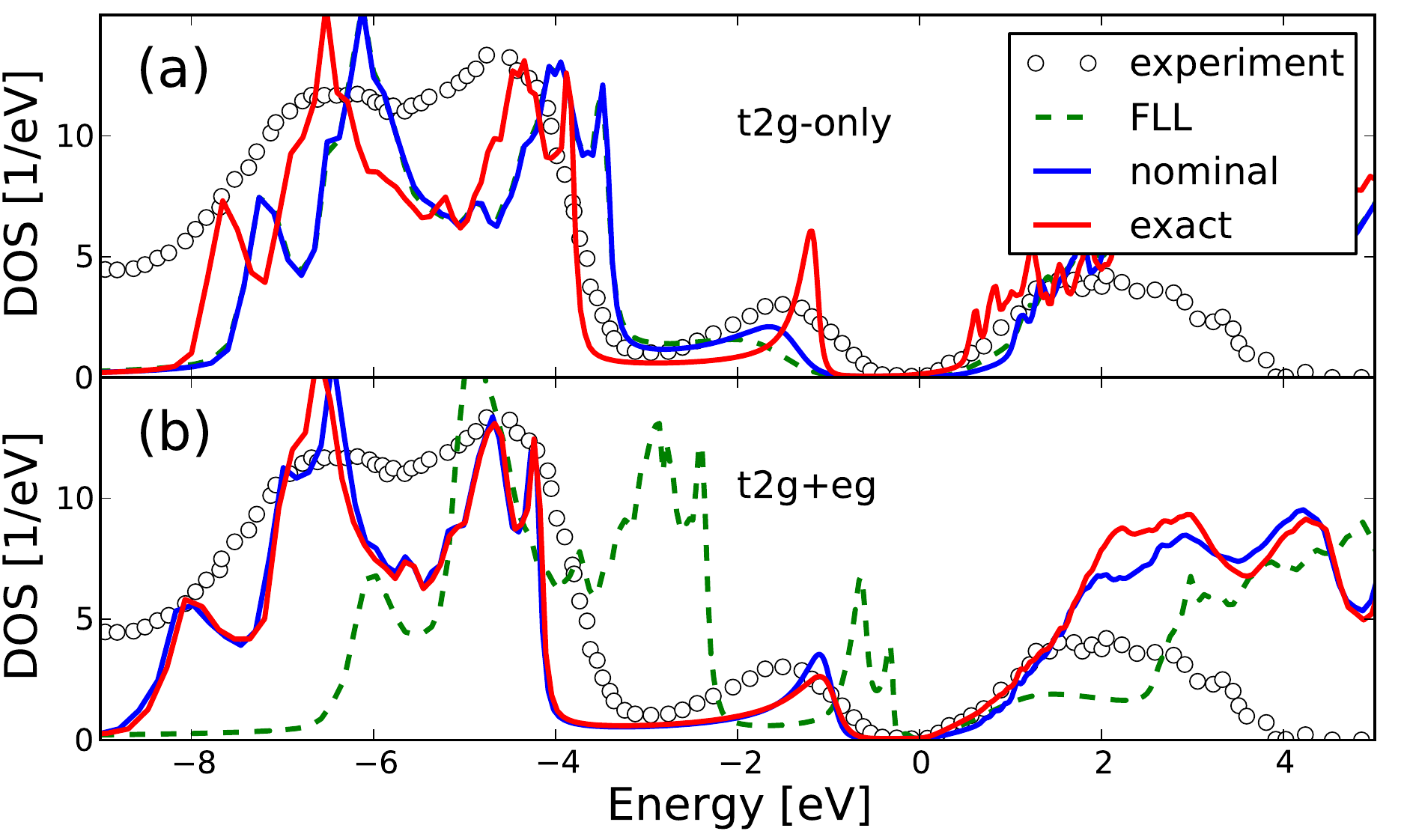}
\caption{(Color online)
LDA+DMFT total density of states for LaVO$_3$ using the three
different DC formulas. (a) only $t2g$ orbitals are treated by DMFT (b)
both $t2g$ and $eg$ orbitals are treated dynamically.
Experimental photoemission is reproduced from Ref.~\onlinecite{LVO-PES}.
}
\label{fig:lavo3}
\end{figure}
Next we present results for the Mott insulating oxide LaVO$_3$, which
is solved in two ways, i) treating only the $t2g$ orbitals dynamically
with DMFT, presented in Table~\ref{Tabel3} and Fig.~\ref{fig:lavo3}a,
and ii) treating both $t2g$ and $eg$ with DMFT. In the first case, the
valences are similar in all three double-counting formulas. The $t2g$
occupancy is very close to nominal value $2$. The exact
double-counting is again smaller than given by FLL or nominal formula,
which leads to a slightly larger splitting between oxygen-p and V-d
states, i.e., slight upward shift of oxygen states in
Fig.~\ref{fig:lavo3}a.  In case ii) displayed in Fig.~\ref{fig:lavo3}b
and tabulated in table~\ref{Tabel4}, where both the $t2g$ and $eg$
orbitals are treated by DMFT, the FLL formula dramatically fails, as
it overestimates the valence, i.e., $n_d^{FLL}-n_d^{exact}\approx
0.26$. While the Mott gap does not entirely collapse, it is severely
underestimated by FLL formula. The nominal valence, however, gives
very similar results as the exact DC. This improvement of nominal DC
as compared to FLL was pointed our in
Refs.~\onlinecite{Haule-DMFT,covalency}, and was found to hold not
just in transition metal oxides but also in
actinides~\cite{BranchingRatios}.
The $t2g$ occupancy $n_{t2g}$ in the nominal and exact DC is very
close to nominal value of 2, equal to the scheme i) presented
above. It is therefore not surprising that the spectra in
Fig. \ref{fig:lavo3}a, and Fig. \ref{fig:lavo3}b are similar, with
slight improvement compared to experiment when $eg$ orbitals are also
treated by DMFT.

In summary, we presented continuum representation of the Dynamical
Mean Field Theory, which allowed us to derive an exact double-counting
between Dynamical Mean Field Theory and Density Functional Theory. The
implementation of exact double-counting for solids shows the improved
agreement with experiment as compared to standard FLL
formula. Previously introduced nominal DC
formula~\cite{Haule-DMFT,covalency} is in very good agreement with
exact double-counting derived here.

This work was supported Simons foundation under project "Many
Electron Problem'', and by NSF-DMR 1405303.

\end{document}